\documentclass[reprint,amsmath,amssymb,aps,superscriptaddress,prx]{revtex4-2}
\usepackage{graphicx}
\usepackage[colorlinks=true,allcolors=blue]{hyperref}
\usepackage{orcidlink}
\usepackage{pifont}
\newcommand{\cmark}{\text{\ding{51}}}
\newcommand{\xmark}{\text{\ding{55}}}

\begin{document}

\title{Saturating the Maximum Success Probability Bound for\texorpdfstring{\\}{ }Noiseless Linear Amplification using Linear Optics}

\author{Joshua J. Guanzon\orcidlink{0000-0002-9990-6341}}
\email{joshua.guanzon@uq.net.au}
\affiliation{Centre for Quantum Computation and Communication Technology, School of Mathematics and Physics, The University of Queensland, St Lucia, Queensland 4072, Australia}

\author{Matthew S. Winnel\orcidlink{0000-0003-3457-4451}}
\affiliation{Centre for Quantum Computation and Communication Technology, School of Mathematics and Physics, The University of Queensland, St Lucia, Queensland 4072, Australia}

\author{Deepesh Singh\orcidlink{0009-0005-9137-2145}}
\affiliation{Centre for Quantum Computation and Communication Technology, School of Mathematics and Physics, The University of Queensland, St Lucia, Queensland 4072, Australia}

\author{Austin P. Lund\orcidlink{0000-0002-1983-3059}}
\affiliation{Dahlem Center for Complex Quantum Systems, Freie Universit\"at Berlin, 14195 Berlin, Germany}
\affiliation{Centre for Quantum Computation and Communication Technology, School of Mathematics and Physics, The University of Queensland, St Lucia, Queensland 4072, Australia}

\author{Timothy C. Ralph\orcidlink{0000-0003-0692-8427}}
\affiliation{Centre for Quantum Computation and Communication Technology, School of Mathematics and Physics, The University of Queensland, St Lucia, Queensland 4072, Australia}

\date{\today}

\begin{abstract}

A noiseless linear amplifier (NLA) performs the highest quality amplification allowable under the rules of quantum physics. Unfortunately, these same rules conspire against us via the no-cloning theorem, which constrains NLA operations to the domain of probabilistic processes. Nevertheless, they are useful for a wide variety of quantum protocols, with numerous proposals assuming access to an optimal NLA device which performs with the maximum possible success probability. Here we propose the first linear optics NLA protocol which asymptotically achieves this success probability bound, by modifying the Knill-Laflamme-Milburn near-deterministic teleporter into an amplifier. 

\end{abstract}

\maketitle

\section{Introduction}

\begin{figure}[t]
    \begin{center}
        \includegraphics[width=\linewidth]{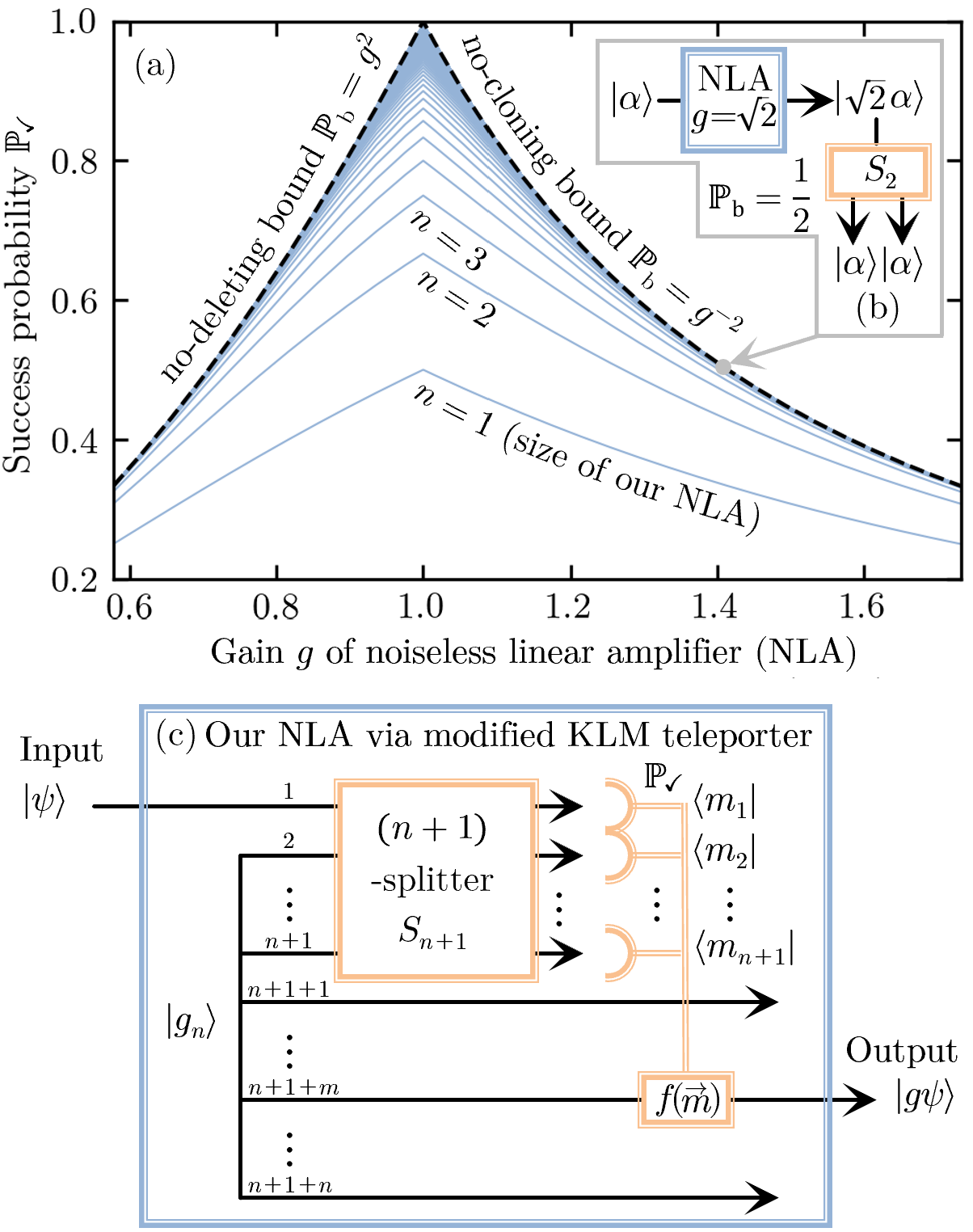}
        \caption{\label{fig:probability} 
            The best quality quantum amplifiers (i.e. NLA) must be probabilistic. They are bounded in probability $\mathbb{P}_\text{b}(g)$ by the dashed black lines in (a), which depends on the amount of amplification (deamplification) gain $g$. This prevents the violation of the no-cloning (no-deleting) theorem on average. The gray inset (b) explains this for $\mathbb{P}_\text{b}(g=\sqrt{2})=1/2$, as this NLA could be used to produce two $|\alpha\rangle$ coherent state clones via a balanced beam-splitter $S_2$. In (c) we propose the first linear optical NLA device, with a size parameter $n$, whose success probability $\mathbb{P}_\cmark$ can saturate the bound. This is Knill-Laflamme-Milburn's near-deterministic teleporter~\cite{knill2001scheme}, but with a weighted entanglement state $|g_n\rangle$ which amplifies the output. We plot $\mathbb{P}_\cmark$ for sequentially increasing $n$ as solid blue lines in (a), assuming the worst possible input state $|\psi\rangle$.}
    \end{center}
\end{figure}

A general definition for an amplifier is a device which increases the amplitude of a signal. The most well-known types are electronic amplifiers, which act on current or voltage; through transistors, they played an important central role in the recent digital technology revolution. A more recent type are quantum amplifiers, which act on quantum states; it remains to be seen whether or not they will play a similar role in the upcoming quantum technology revolution. In this respect, it is important to properly investigate what is actually physically achievable using quantum amplifiers.

In this paper we will consider quantum amplifiers that perform the best quality amplification. To understand the properties that such an amplifier would satisfy, consider a scenario where we want to amplify a coherent state $|\alpha\rangle$ with $\alpha$ complex amplitude. Recall that coherent states have minimum noise profiles according to the uncertainty principle. A noiseless amplifier with gain $g\in(0,\infty)$ should produce another coherent state $|g\alpha\rangle$ since it introduces no extra noise. Furthermore, a noiseless linear amplifier (NLA) can do this without any prior knowledge of $\alpha$; in other words, it acts like the operator $g^{a^\dagger a}$, since $g^{a^\dagger a}|\alpha\rangle \propto |g\alpha\rangle$. 

This NLA operation is so powerful that it was believed to be impossible in the past~\cite{heffner1962fundamental}. However, it is possible to make probabilistic NLAs~\cite{ralph2009nondeterministic}, with a success probability bound $\mathbb{P}_\text{b}$ shown by the dashed black lines in Fig.~\ref{fig:probability}(a). Intuitively, this is because using $|g\alpha\rangle$ with $g\geq 1$ and a balanced beam-splitter one can produce up to $g^2$ clones of an unknown $|\alpha\rangle$~\cite{fiuravsek2004optimal}, as shown for $g=\sqrt{2}$ in Fig.~\ref{fig:probability}(b). Therefore, to ensure that the no-cloning theorem~\cite{wootters1982single,dieks1982communication} isn't violated we require no extra clones are produced on average $g^2\mathbb{P}_\text{b} = 1$; this gives us the no-cloning bound $\mathbb{P}_\text{b}(g\geq1) = g^{-2}$~\cite{mcmahon2014optimal}. One can apply the reverse logic and the no-deleting theorem~\cite{kumar2000impossibility} to produce the bound $\mathbb{P}_\text{b}(g\leq1) = g^2$ for deamplification. Note that there exists more formal methods of deriving this bound, such as via quantum state discrimination~\cite{pandey2013quantum}.   

Despite NLAs being nondeterministic, the unrivaled quality of noiseless amplification means they are often the only path forward for many quantum protocols. This includes for applications in quantum communication~\cite{gisin2010proposal,blandino2012improving,mivcuda2012noiseless,xu2013improving,ghalaii2020long,zhou2020device,li2021improving,xu2021feasible,villasenor2021enhancing,he2022teleportation,zhao2023enhancing,notarnicola2023long}, quantum repeater networks~\cite{dias2017quantum,dias2020quantum,seshadreesan2020continuous,winnel2021overcoming,laurenza2022rate,tillman2022supporting,dias2022distributing,ghalaii2022composable,bjerrum2023quantum}, quantum entanglement distillation~\cite{zhang2012protecting,seshadreesan2019continuous,liu2022distillation,he2022teleportation2,mauron2022comparison}, quantum improved sensing~\cite{usuga2010noise,zhao2017quantum,xia2019repeater,karsa_ghalaii_pirandola_2022,tang2023improving}, and quantum error correction~\cite{ralph2011quantum,dias2018quantum,slussarenko2022quantum}. These protocols either presuppose the use of, or otherwise can be enhanced by, a maximally efficient NLA with a success probability equivalent to the bound. However, it is not apparent how to even implement such an efficient NLA in optics, the natural platform for many of these schemes, without strong non-linear interactions. It was shown in Ref.~\cite{mcmahon2014optimal} that a maximally efficient NLA could be constructed if we could somehow non-destructively interact the input light with a qubit system, which requires large experimental overheads. It also claimed in Ref.~\cite{mcmahon2014optimal} that for all known linear optical NLAs the success probability is only $(1+g^2)^{-1}$. We will prove that linear optical interactions can achieve the success probability bound. 

There are various methods which can perform the NLA operation~\cite{xiang2010heralded,fiuravsek2009engineering,zavatta2011high,ferreyrol2010implementation,jeffers2010nondeterministic,marek2010coherent,neergaard2013quantum,zhao2017characterization,zhang2018photon,hu2019entanglement,he2021noiseless,zhao2022entanglement}, with sub-optimal success probability. In particular, there has been a recent resurgence of research on NLAs~\cite{winnel2020generalized,guanzon2022ideal,fiuravsek2022teleportation,fiuravsek2022optimal,zhong2022quantum,guanzon2023noiseless,goldberg2023teleamplification} that work via quantum teleportation~\cite{bennett1993teleporting}, which are called teleamplifiers. We highlight this type of NLAs because if we want a teleamplifier that can saturate the probability bound, then clearly it should also be able to act like a deterministic teleporter since $\mathbb{P}_\text{b}(g=1)=1$. Fortunately, there already exists a deterministic teleporter as part of the well-known Knill-Laflamme-Milburn (KLM) linear optical quantum computing protocol~\cite{knill2001scheme}. This inspires the simple idea behind this work: modify KLM's teleporter to operate as a teleamplifier, and then verify that it is the first linear optical NLA proposal which can actually saturate the maximum probability bound $\mathbb{P}_\text{b}$ for all $g$. 

We begin in Section~\ref{sec:proof}, where we describe our scalable teleamplifier and prove that it operates as a NLA. We then calculate the success probability of our teleamplifier in Section~\ref{sec:prob}, and show that it asymptotically saturates the probability bound. In Section~\ref{sec:exp}, we investigate the experimental resource requirements of our protocol at the smallest sizes. We also describe how to extend the results to multiphoton input states in Section~\ref{sec:multi}. Finally, we conclude in Section~\ref{sec:con}.

\section{Our Noiseless Linear Amplifier} \label{sec:proof}

Suppose we start with an unknown quantum input state $|\psi\rangle$ containing up to a single photon (i.e. a single-rail qubit), then a NLA operation $g^{a^\dagger a}$ should result in the following output 
\begin{align}
    |\psi\rangle = c_0 |0\rangle + c_1 |1\rangle \rightarrow |g\psi\rangle = c_0 |0\rangle + g c_1 |1\rangle, \label{eq:NLA}
\end{align}
where $g\in(0,\infty)$ is the amount of gain. We propose a scalable linear optical NLA protocol, with a size parameter $n\in\mathbb{N}$, as shown in Fig.~\ref{fig:probability}(c). We will later show it can saturate the maximum success probability bound at asymptotically large sizes $n\to\infty$. However, we will firstly verify that our device actually performs the claimed NLA operation in Eq.~\eqref{eq:NLA} and produces the correct amplified output state. 

Our protocol requires the following $n$ single-photons entangled over $2n$ modes resource state
\begin{align}
    |g_n\rangle = \frac{1}{\sqrt{\mathcal{N}}} \sum^n_{j=0} g^{n-j} |1\rangle^j |0\rangle^{n-j} |0\rangle^j |1\rangle^{n-j}, \label{eq:res}
\end{align}
with a normalisation factor $\mathcal{N} = \sum_{j=0}^n g^{2j}$. We use the notation $|s\rangle^k = \otimes^k_{j=1}|s\rangle$ to mean that there are $k$ modes each occupied with $s$ photons. Note for $g=1$ this $|g_n\rangle$ state reduces down to the entanglement resource state used in KLM's near-deterministic teleporter~\cite{knill2001scheme}. We also require an $(n+1)$-splitter with a scattering matrix
\begin{align}
    (S_{n+1})_{j,k} \equiv \frac{\omega^{(j-1)(k-1)}_{n+1}}{\sqrt{n+1}},\quad  \omega_{n+1}\equiv e^{-i2\pi/(n+1)},
\end{align}
whose phase $\omega_{n+1}$ configuration follows the quantum Fourier transformation. This operation scatters photons linearly as $(a_1^\dagger,\ldots,a_{n+1}^\dagger)^T \rightarrow S_{n+1} (a_1^\dagger,\ldots,a_{n+1}^\dagger)^T$. Hence this $(n+1)$-splitter could be thought of as an $(n+1)$ mode generalisation of a balanced beam-splitter. 

As shown in Fig.~\ref{fig:probability}(c), the first step of our protocol is to mix the unknown input and the first $n$ modes of the resource state on the $(n+1)$-splitter as
\begin{align}
    S_{n+1}|\psi\rangle|g_n\rangle &= \sum^n_{j=0} \frac{g^{n-j}}{\sqrt{\mathcal{N}}} [c_0 S_{n+1}|0\rangle |1\rangle^j |0\rangle^{n-j} \nonumber \\ 
    &\quad + c_1 S_{n+1}|1\rangle |1\rangle^j |0\rangle^{n-j} ] |0\rangle^j |1\rangle^{n-j}. \label{eq:mixing}
\end{align}
There are $2(n+1)$ terms in this summation, but recall we ultimately want just a pair of terms (i.e. two terms) of the form $c_0|0\rangle+gc_1|1\rangle$. We select particular terms by performing measurements which give us the total number of photons that exit the $(n+1)$-splitter $m$. Notice that the $c_0$ terms $S_{n+1}|0\rangle |1\rangle^j |0\rangle^{n-j}$ has $m=j$ photons, while the $c_1$ terms $S_{n+1}|1\rangle |1\rangle^j |0\rangle^{n-j}$ has $m=j+1$ photons, where $j\in\{0,\ldots,n\}$. Therefore, one can correspond a pair of terms for $m\in\{1,\ldots,n\}$ photons exiting the $(n+1)$-splitter. There is also a single $(c_0,j=0)$ term with $m=0$ no photons, and a single $(c_1,j=n)$ term with $m=n+1$ photons; selecting on these $m$ values will produce a single state output which is an error. We will now verify that the rest of the $m$ values gives the correct amplified output state. 

We perform photon number measurements $\langle \vec{m}| \equiv \langle m_1 | \langle m_2 | \cdots \langle m_{n+1}|$ right after the $(n+1)$-splitter, in which we measured $m = \sum_{i=1}^{n+1} m_i$ total photons. Based on our previous discussion, it is clear that this outcome must be due to either the $(c_0,j=m)$ term or the $(c_1,j=m-1)$ term, as all other terms do not have the correct number of photons. These two terms have non-zero probability amplitudes and, as discussed in the supplementary of KLM~\cite{knill2001scheme}, due to the symmetry of $S_{n+1}$ they are related as follows
\begin{align}
    \langle \vec{m}|S_{n+1}|1\rangle^m |0\rangle^{n-m+1} &\equiv p, \\ 
    \langle \vec{m}|S_{n+1}|0\rangle |1\rangle^m |0\rangle^{n-m} &= \omega^{f(\vec{m})}_{n+1} p,
\end{align}
which we formally prove in Appendix~\ref{sec:permproof}. In other words, these probability amplitudes differ only by a correctable phase which depends on the known measurement outcome $f(\vec{m})=\sum_{k=1}^{n+1} (k-1)m_k$. Using these results and Eq.~\eqref{eq:mixing}, the output state will be
\begin{align}
    &\langle \vec{m}|S_{n+1}|\psi\rangle|g_n\rangle = \frac{g^{n-m}}{\sqrt{\mathcal{N}}} c_0  \omega^{f(\vec{m})}_{n+1} p |0\rangle^m |1\rangle^{n-m} \nonumber \\ 
    &\hspace{80pt} + \frac{g^{n-m+1}}{\sqrt{\mathcal{N}}} c_1 p |0\rangle^{m-1} |1\rangle^{n-m+1} \nonumber \\ 
    &= \frac{g^{n-m}p}{\sqrt{\mathcal{N}}} |0\rangle^{m-1} [   \omega^{f(\vec{m})}_{n+1} c_0 |0\rangle + g c_1 |1\rangle ] |1\rangle^{n-m}. 
\end{align}
Finally, by applying a simple phase correction $\omega_{n+1}^{f(\vec{m})a^\dagger_m a_m}$ to the $m$th output mode, we get the final output state
\begin{align}
    &\omega_{n+1}^{f(\vec{m})a^\dagger_m a_m} \langle \vec{m}|S_{n+1}|\psi\rangle|g_n\rangle \nonumber \\ 
    &= \frac{\omega^{f(\vec{m})}_{n+1} g^{n-m}p}{\sqrt{\mathcal{N}}}  |0\rangle^{m-1} [ c_0 |0\rangle + g c_1 |1\rangle ] |1\rangle^{n-m}. \label{eq:output}
\end{align}
Thus we have verified that on the $m$th output mode we get the required NLA output state $|g\psi\rangle = c_0 |0\rangle + g c_1 |1\rangle$, for any $\langle \vec{m}|$ given we measured $m\in\{1,\ldots,n\}$ total photons. By tracing over all output modes but the $m$th mode and renormalizing, we obtain $|g \psi \rangle$ as expected.

\section{Success Probability Analysis} \label{sec:prob}

Let us now determine the success probability for this NLA protocol $\mathbb{P}_\cmark$. We could calculate this by using the output state given in Eq.~\eqref{eq:output} and summing over all relevant success measurements $\{\forall \vec{m}|m\in\{1,\ldots,n\}\}$. However, it is easier to infer the success probability $\mathbb{P}_\cmark = 1 - \mathbb{P}_\xmark$ from the failure probability, since there are just two failure cases $\mathbb{P}_\xmark = \mathbb{P}_{m=0} + \mathbb{P}_{m=n+1}$. Recall that the $m=0$ case with $\langle\vec{m}| = \langle 0|^{n+1}$ only has non-zero overlap with the isolated $(c_0,j=0)$ term in Eq.~\eqref{eq:mixing}, which gives the output
\begin{align}
    \langle 0|^{n+1}S_{n+1}|\psi\rangle|g_n\rangle &= \frac{g^nc_0}{\sqrt{\mathcal{N}}} |1\rangle^n.
\end{align}
Thus the probability of measuring $m=0$ is 
\begin{align}
    \mathbb{P}_{m=0} = |\langle 0|^{n+1}S_{n+1}|\psi\rangle|g_n\rangle|^2 
    = \frac{g^{2n}|c_0|^2}{\mathcal{N}}. 
\end{align}
Likewise, the $m=n+1$ case with $\langle \vec{m}_{n+1}|$  where $\vec{m}_{n+1}\in \{\forall \vec{m}|m=n+1\}$ only has non-zero overlap with the isolated $(c_1,j=n)$ term in Eq.~\eqref{eq:mixing}, which gives the output
\begin{align}
    \langle \vec{m}_{n+1}|S_{n+1}|\psi\rangle|g_n\rangle &= \frac{c_1}{\sqrt{\mathcal{N}}} \langle \vec{m}_{n+1}|S_{n+1}|1\rangle^{n+1} |0\rangle^n.
\end{align}
Thus the probability of measuring $m=n+1$ is 
\begin{align}
    \mathbb{P}_{m=n+1} &= \sum_{\vec{m}|m=n+1} |\langle \vec{m}|S_{n+1}|\psi\rangle|g_n\rangle|^2 \nonumber \\  
    &= \frac{|c_1|^2}{\mathcal{N}} \sum_{\vec{m}|m=n+1} \langle 1|^{n+1} S_{n+1}^\dagger | \vec{m} \rangle \langle \vec{m} |S_{n+1}|1\rangle^{n+1} \nonumber \\  
    &= \frac{|c_1|^2}{\mathcal{N}}. 
\end{align}
Note the explanation for the third equality is that $\sum_{\vec{m}|m=n+1} | \vec{m} \rangle \langle \vec{m} |$ is the identity operator for states made from $n+1$ photons in $n+1$ modes. For example, the $n=1$ operator $\sum_{\vec{m}|m=2} | \vec{m} \rangle \langle \vec{m} | = |0\rangle|2\rangle\langle0|\langle2| + |1\rangle|1\rangle\langle1|\langle1| + |2\rangle|0\rangle\langle2|\langle0|$ acting on any $2$ photons in $2$ modes state $c_{02}|0\rangle|2\rangle+c_{11}|1\rangle|1\rangle+c_{20}|2\rangle|0\rangle$ will leave it unchanged. Since $S_{n+1}|1\rangle^{n+1}$ is an $n+1$ photons in $n+1$ modes state, it will remain unchanged by the operator $\sum_{\vec{m}|m=n+1} | \vec{m} \rangle \langle \vec{m} |$. 

By considering the complement of these failure outcomes, we can calculate the success probability to be
\begin{align}
    \mathbb{P}_\cmark &= 1 - (\mathbb{P}_{m=0} + \mathbb{P}_{m=n+1}) \nonumber \\
    &= 1 - \frac{g^{2n}|c_0|^2 + |c_1|^2 }{\mathcal{N}}. 
\end{align}
The closed expression for the normalisation factor $\mathcal{N}$ is given by the geometric series
\begin{align}
    \mathcal{N} = \sum_{j=0}^n g^{2j} = \begin{cases}
        \frac{1-g^{2(n+1)}}{1-g^2}, & g \neq 1, \\ 
        n+1, & g = 1. 
    \end{cases} 
\end{align}
Using this expression and the input normalisation condition $|c_0|^2+|c_1|^2=1$, we can get a simple closed expression for the success probability as
\begin{align}
    \mathbb{P}_\cmark &= \begin{cases}
        \frac{(1-g^{2n})(|c_0|^2+g^2|c_1|^2)}{1-g^{2(n+1)}}, & g \neq 1, \\ 
        \frac{n}{n+1}, & g = 1. 
    \end{cases}
\end{align}
As expected for $g=1$, this $n/(n+1)$ expression matches the success probability for KLM's near-deterministic teleporter~\cite{knill2001scheme}. We plot this success probability $\mathbb{P}_\cmark$ in Fig.~\ref{fig:probability}(a) assuming the worst possible input state; in other words, essentially vacuum $|c_0|^2\simeq1$ for the amplification region $g>1$, and essentially single photon $|c_0|^2\simeq0$ for the deamplification region $g<1$. Note we are saying `essentially' because one can't amplify a single number state. The solid blue lines in Fig.~\ref{fig:probability}(a) shows that the success probability will always improve as we increase the size $n$ of our teleamplifier. By the derivative test, one can also analytically verify that the factor $(1-g^{2n})/(1-g^{2(n+1)})$, and hence $\mathbb{P}_\cmark$, always increases with $n$ for all $g$. 

If we consider the asymptotic limit of $n\to\infty$, we get the following expression for the factor
\begin{align}
    \lim_{n\to\infty} \frac{1-g^{2n}}{1-g^{2(n+1)}} &= \begin{cases}
        g^{-2}, & g > 1, \\ 
        1, & g < 1.  
    \end{cases} 
\end{align}
Hence the maximum achievable success probability of our teleamplifier depends on whether we are considering amplification $g>1$, teleportation $g=1$, or deamplification $g<1$, as summarised below  
\begin{align}
    \lim_{n\to\infty} \mathbb{P}_\cmark &= \begin{cases}
        g^{-2}|c_0|^2 + |c_1|^2, & g > 1, \\ 
        1, & g = 1, \\ 
        |c_0|^2 + g^2|c_1|^2, & g < 1.
    \end{cases}
\end{align}
This is a very nice expression, as we can see how the gain $g$ affects the success probability through the different input components. At $g=1$, we have the deterministic teleporter, and if we move away from $g=1$ we generally have a reduction in success probability. In the amplification regime $g>1$, this reduction acts on the $|c_0|^2$ vacuum component in the input. This makes sense, because if $|c_0|^2\simeq0$ then our input is essentially already completely amplified $|1\rangle$, hence this should be deterministic. We note that even in the worst case scenario with an input of $|c_0|^2\simeq1$, our teleamplifier's success probability saturates the no-cloning bound $\mathbb{P}_\text{b}(g\geq1) = g^{-2}$. This same analysis can be applied in reverse for the deamplification regime, which shows our teleamplifier can also achieve the no-deleting bound $\mathbb{P}_\text{b}(g\leq1) = g^2$.

\section{Experimental Considerations} \label{sec:exp}

What we demonstrated has theoretically significance; that mere linear interactions, with entanglement and measurements, are sufficient to perform amplification with the highest quality and efficiency allowed by quantum physics. Let us now consider the experimental requirements of our protocol in practice. 

One major source of experimental complexity is the required entangled resource state $|g_n\rangle$, as defined in Eq.~\eqref{eq:res}. The smallest size $|g_1\rangle$ can be made deterministically using a single photon and beam-splitter; in fact, this is equivalent to the single-photon quantum scissor NLA~\cite{ralph2009nondeterministic}. However, $|g_2\rangle$ can't be made just from two single photons and a linear optical network; it must be prepared offline before the unknown input state arrives. One suggestion~\cite{su2019conversion,quesada2019simulating} is to use a Gaussian Boson sampling-like device~\cite{hamilton2017gaussian}, which we optimised via machine learning~\cite{sabapathy2019production} to produce $|g_2\rangle$ with very high fidelity $F>0.999$. This is a shotgun approach with low efficiencies, instead we can draw inspiration from tailored approaches from past studies on KLM's resource state~\cite{myers2007investigating}. In particular, we propose a construction method like Ref.~\cite{franson2004generation} for all $|g_n\rangle$ using $n-1$ controlled beam-splitters, which can be implemented with linear optics tools with $1/16^{n-1}$ post-selection probability~\cite{myers2007investigating}. The details are in Appendix~\ref{sec:gen}.

\begin{figure}[t]
    \begin{center}
        \includegraphics[width=\linewidth]{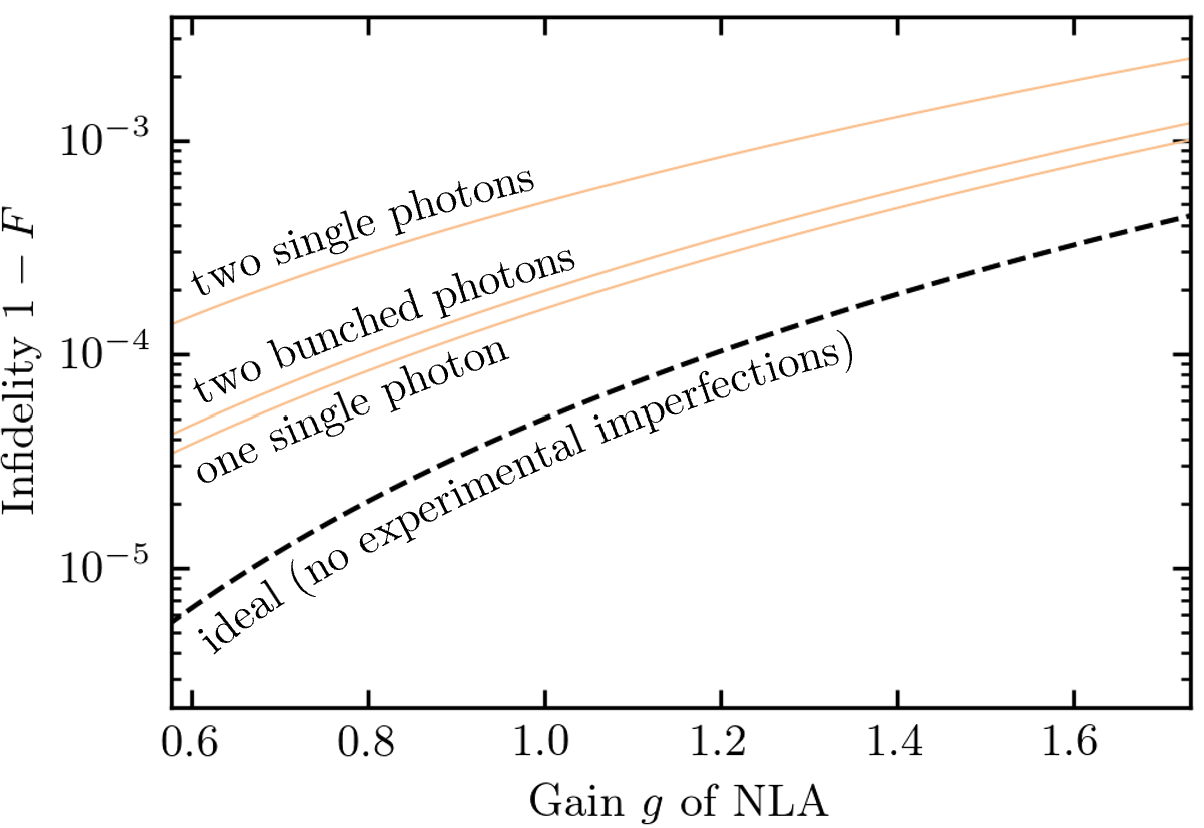}
        \caption{\label{fig:fidelity} 
            The fidelity of our protocol appears to depend on the type of detection $\langle \vec{m}|$, rather than size $n$. Here we consider amplifying a coherent state $|\alpha\rangle$ with $\alpha=0.1$, the dashed line is due to single-photon cut-off. Imperfections are pure loss with transmissivity $\eta = 0.7$, applied independently before all detectors and on all resource modes.}
    \end{center}
\end{figure}

Another major consideration is experimental error. The $(\forall n, g=1)$ edge case is KLM's teleporter, which is known to be fault-tolerant from photon loss and detector inefficiencies. Furthermore, the $(n=1,\forall g)$ edge case is the single-photon quantum scissor NLA, which has been experimentally implement~\cite{xiang2010heralded} and is also known to be loss-tolerant~\cite{winnel2021overcoming}. Therefore, one would expect that larger $n$ is also loss-tolerant, which we verified via numerical simulation of $n\in\{1,2\}$. We show in Fig.~\ref{fig:fidelity} that the most likely outcome of a single photon detection gives equivalent fidelity, while the second most likely outcome of a two bunched photons detection also gives close fidelity. Note also that the probability of detecting a single photon increases as the $g$ gain increases. Thus our proposal can work well right out of the box, and also has the possibility of improvement by leveraging existing error correction protocols made for KLM~\cite{hayes2004utilizing}. 

The predominate type of error in optics is loss, which especially affects multiphoton states. This is why numerous quantum optical protocols rely on just single-photon qubit-like states. For example, Ref.~\cite{winnel2021overcoming} showed that whilst a multiphoton NLA could distill a larger magnitude of entanglement through a lossy channel, a single-photon NLA has much better entanglement rates (which takes into account success probability). Therefore, our current proposed NLA with single-photon output will be useful for achieving the best possible entanglement rates and key rates for quantum communication purposes.

\section{Extension to Multiphoton States} \label{sec:multi}

\begin{figure}[t]
    \begin{center}
        \includegraphics[width=\linewidth]{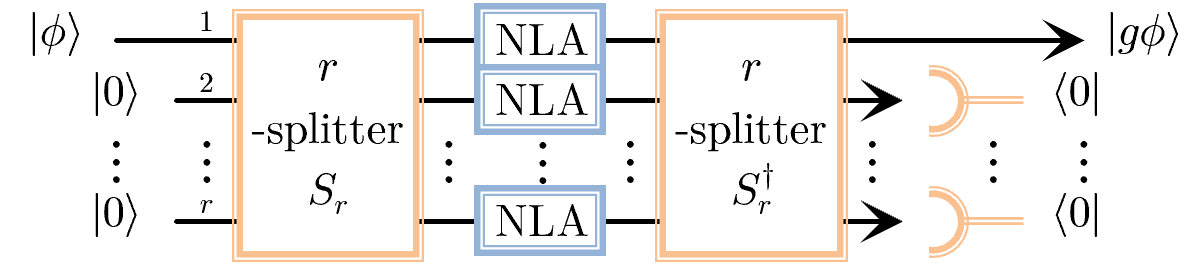}
        \caption{\label{fig:multiphoton} 
            Our results can be extended to unknown input states containing multiple photons $|\phi\rangle$. Intuitively, the input light is split by $S_r$ into $r$ rails, such that each rail has less than one photon on average. Each rail goes through our proposed NLA, and then recombined using $S_r^\dagger$ into a single output rail.}
    \end{center}
\end{figure}

There are instances where we may want to amplify unknown quantum states of light that contain multiple photons $|\phi\rangle = \sum_{j=0}^\infty c_j |j\rangle$. Any physical state should be bounded in energy, where $r\gg \langle \phi | a^\dagger a | \phi \rangle$. Therefore, if we split up $|\phi\rangle$ using a large enough $r$-splitter $S_r$ as shown in Fig.~\ref{fig:multiphoton}, we can always manufacture a situation where each rail has much less than one photon on average. We can then apply our NLA to amplify each rail almost ideally. Finally, we apply the inverse $r$-splitter $S_r^\dagger$, which should recombine the light back into a single rail $|g\phi\rangle$. Therefore, the success probability of this device is at most $\lim_{n\to\infty}\mathbb{P}_\cmark^r=g^{-2r}$ for $g\geq1$. This beats the $(1+g^2)^{-r}$ success probability for other linear optical designs~\cite{mcmahon2014optimal}. 

This extension method was proposed in Ref.~\cite{xiang2010heralded} using the single-photon quantum scissor NLA. However, it can be used to extend any single-photon NLA device into accepting multiple photons, since they perform the same operation. We must emphasise that this extension only approximately performs $g^{a^\dagger a}$ on the multiphoton terms. This is because the output $|g\phi\rangle = N \sum_{j=0}^r d_j g^j c_j |j\rangle$ has distortions $d_j$, which should be minor (i.e. $d_j\approx 1$ for small $j$) if $r$ is large enough. Note there is a recently discovered method for teleamplifying multiphoton states with perfect fidelity (i.e. without these distortions $d_j=1$ for all $j$) in Ref.~\cite{guanzon2022ideal}, however this isn't done at the multiphoton success probability bound $\mathbb{P}_{\text{b},r}=g^{-2r}$~\cite{mcmahon2014optimal}. We will leave as an open question whether it is possible to combine Ref.~\cite{guanzon2022ideal} with our current proposal to construct a maximally efficient, perfect fidelity multiphoton NLA.

\section{Conclusion} \label{sec:con}

We have presented a scalable method of constructing a quantum amplifier which performs the best quality amplification, using only linear optical tools. We verified that it is the first amplifier of it's kind which saturates the quantum limited bounds of success probability. There are numerous quantum optics protocols which assume access to such a maximally efficient amplifier. Therefore, our proposal provides a clear methodology to do this purely in optics, without the experimental difficulties of switching platforms or strong non-linearities. Instead, our method relies on an entangled resource state, which we showed can be prepared beforehand with high fidelity and also using just linear optical tools. Finally, we demonstrated how we can extend our proposed method to allow multiphoton inputs, with much better success probabilities than any other linear optical amplifier. 

It is easy to conceive situations where the input quantum signal is very important or costly, and we want to do all we can to maximise our chances of correctly amplifying it. This is especially the case if we are considering using amplifiers in each node of a large quantum network (i.e. the quantum internet), whose success probabilities will multiply together. In these situations, it is likely that the benefits of our proposal outweigh the experimental costs of implementing it. These costs are also rapidly decreasing as of late due to large investments in quantum optical protocols, which has made improvements in integrated linear optical circuits, photon detectors, and generating photonic entangled cluster states. 

More broadly, this research reveals that experimentally challenging nonlinear optical effects can be unnecessary. Instead linear optical effects, with entanglement and photon detections, are sufficient to perform useful operations even at the maximum limits allowed by quantum physics. In this regard, it is fitting that a tool from the KLM scheme, the first linear optical universal quantum computing protocol, provided the necessary scaffolding to reveal this for quantum amplification.

\begin{acknowledgments}
This research was supported by the Australian Research Council Centre of Excellence for Quantum Computation and Communication Technology (Project No. CE170100012).
\end{acknowledgments}

\appendix

\section{Amplitude using Permanents Proof} \label{sec:permproof}

In this section, we will prove the following relation
\begin{align}
    \langle \vec{m}|S_{n+1}|1\rangle^m |0\rangle^{n-m+1} &\equiv p, \\ 
    \langle \vec{m}|S_{n+1}|0\rangle |1\rangle^m |0\rangle^{n-m} &= \omega^{\sum_{k=1}^{n+1} (k-1)m_k}_{n+1} p. \label{eq:toprove}
\end{align}
In other words, we want to verify that these two probability amplitudes are related by just a correctable phase shift, for any given photon number measurement outcome $\langle \vec{m}| \equiv \langle m_1 | \langle m_2 | \cdots \langle m_{n+1}|$ with $\sum_{i=1}^{n+1}m_i=m$ total photons. As an aside, we will also show that the probability amplitude is given by
\begin{align}
    p = \frac{\text{Per}( \Omega_{\vec{m}})}{\sqrt{\prod_{i=1}^{n+1}m_i!}}. 
\end{align}
Here $\text{Per}$ is the permanent matrix function, which is calculated like the determinant but without the alternating negative factor. 

The $(n+1)$-splitter splits a single beam into $(n+1)$ equal beams. This operation is therefore associated with the following unitary matrix 
\begin{align}
    S_{n+1} = \frac{1}{\sqrt{n+1}}
        \overbrace{\begin{bmatrix}
            1 & 1 & \cdots & 1 \\ 
            1 & \omega_{n+1} & \cdots & \omega^n_{n+1} \\
            \vdots & \vdots & \ddots & \vdots  \\ 
            1 & \omega^n_{n+1} & \cdots & \omega^{2n}_{n+1} 
        \end{bmatrix} }^{n+1 \text{ columns}} 
        \left. \phantom{\begin{bmatrix}
            1 \\ 
            1 \\
            \vdots \\ 
            1 
        \end{bmatrix}\hspace{-2.3em}} \right\}{\scriptstyle n+1 \text{ rows}}, 
\end{align}
which describes the scattering of photons as follows $(a_1^\dagger,\ldots,a_{n+1}^\dagger)^T \rightarrow S_{n+1}(a_1^\dagger,\ldots,a_{n+1}^\dagger)^T$. For example, injecting a single photon into the $c$th input port $a_c^\dagger |0\rangle^{n+1}$ is related to $c$th column of $S_{n+1}$, where $a_c^\dagger \rightarrow \sum_{r=1}^{n+1} (S_{n+1})_{r,c} a_r^\dagger = \sum_{r=1}^{n+1} \frac{\omega^{(c-1)(r-1)}_{n+1}}{\sqrt{n+1}} a_r^\dagger$. Notice that the probability amplitude for a particular single-photon input and output combination is 
\begin{align}
    \langle 0|^{n+1} a_r S_{n+1} a_c^\dagger |0\rangle^{n+1} = \frac{\omega^{(c-1)(r-1)}_{n+1}}{\sqrt{n+1}}, 
\end{align}
which is the $(r,c)$ element in $S_{n+1}$. For a more general multiple photons input, the scattering probability amplitudes are related to multiple elements of $S_{n+1}$. Suppose we have an arbitrary $m$ photons input state $|\vec{m}'\rangle \equiv \otimes_{i=1}^{n+1} | m'_i \rangle = \prod_{i=1}^{n+1}[(a_i^\dagger)^{m'_i}/\sqrt{m'_i!}]|0\rangle^{n+1}$, where $\sum_{i=1}^{n+1}m'_i=m$, then we have the following probability amplitude 
\begin{align}
    \langle \vec{m}| S_{n+1} |\vec{m}'\rangle = \frac{\text{Per}(\Omega_{\vec{m},\vec{m}'})}{\sqrt{\prod_{i=1}^{n+1}m_i! m'_i!}}. \label{eq:Aper}
\end{align}
This is a well-known relation from boson sampling studies~\cite{scheel2004permanents,gard2015introduction,lund2017quantum}, for example Eq.~(7) in Ref.~\cite{lund2017quantum}. The $\Omega_{\vec{m},\vec{m}'}$ is an $m \times m$ matrix constructed by taking the elements of $S_{n+1}$ by the repeating column $c$ $m'_c$ times and row $r$ $m_r$ times. This is more easily understood by considering a specific example, as outlined in the next paragraph. 

Let us consider the $n=2$ amplifier case. This requires the $3$-splitter, associated with the following scattering unitary matrix
\begin{align}
    S_3 = \frac{1}{\sqrt{3}}
        \begin{bmatrix}
            1 & 1 & 1 \\ 
            1 & \omega_3 & \omega^2_3 \\
            1 & \omega^2_3 & \omega^4_3 
        \end{bmatrix}.
\end{align}
Suppose we want to calculate the probability amplitude for the input $\vec{m}'=(1,1,0)$ and output $\vec{m}=(0,2,0)$. This means we need to consider the elements of $S_3$ made from the first two columns and repeating the second row twice, as follows
\begin{align}
    \Omega_{(0,2,0),(1,1,0)} = \frac{1}{\sqrt{3}} 
        \begin{bmatrix}
            1 & \omega_3 \\
            1 & \omega_3
        \end{bmatrix}.
\end{align}
Taking the permanent of this matrix, we can now calculate the probability amplitude as
\begin{align}
    p' = \langle0|\langle2|\langle0|S_3|1\rangle|1\rangle|0\rangle = \frac{\text{Per}(\Omega_{(0,2,0),(1,1,0)})}{\sqrt{2}},
\end{align}
using Eq.~\eqref{eq:Aper}. Similarly for a different input $\vec{m}'=(0,1,1)$ but same output $\vec{m}=(0,2,0)$, we construct the following matrix 
\begin{align}
    \Omega_{(0,2,0),(0,1,1)} = \frac{1}{\sqrt{3}} 
        \begin{bmatrix}
            \omega_3 & \omega^2_3 \\
            \omega_3 & \omega^2_3
        \end{bmatrix}, 
\end{align}
whose permanent is used to calculate the probability amplitude $\langle0|\langle2|\langle0|S_3|0\rangle|1\rangle|1\rangle$. However, there is a property of permanents where 
\begin{align}
    \text{Per} \begin{bmatrix}
            a u_{11} & a u_{12} \\
            b u_{21} & b u_{22} 
        \end{bmatrix} 
    = ab \text{Per} \begin{bmatrix}
            u_{11} & u_{12} \\
            u_{21} & u_{22} 
        \end{bmatrix}, \label{eq:perrule} 
\end{align}
which we can use to connect the two permanents under investigation
\begin{align}
    \text{Per}(\Omega_{(0,2,0),(0,1,1)}) &= \text{Per}\left( \frac{1}{\sqrt{3}} 
        \begin{bmatrix}
            \omega_3 & \omega^2_3 \\
            \omega_3 & \omega^2_3
        \end{bmatrix} \right) \nonumber \\ 
    &= \omega_3^2 \text{Per}\left( \frac{1}{\sqrt{3}} 
        \begin{bmatrix}
            1 & \omega_3 \\
            1 & \omega_3
        \end{bmatrix} \right) \nonumber \\ 
    &= \omega_3^2 \text{Per}(\Omega_{(0,2,0),(1,1,0)}).  
\end{align}
Thus we have shown that these two amplitudes are related by just a phase
\begin{align}
    \langle0|\langle2|\langle0|S_3|0\rangle|1\rangle|1\rangle &= \omega_3^2 \langle0|\langle2|\langle0|S_3|1\rangle|1\rangle|0\rangle \nonumber \\ 
    &= \omega_3^2p', 
\end{align}
which agrees with Eq.~\eqref{eq:toprove}, since $\sum_{k=1}^{n+1}(k-1)m_k = 2$ for this $\vec{m}=(0,2,0)$ example.  

We consider generalising this result to more modes, with inputs of the form $|1\rangle^m |0\rangle^{n-m+1}$ and $|0\rangle |1\rangle^m |0\rangle^{n-m}$, and arbitrary measurements $\langle \vec{m}|$. For the amplitude $\langle \vec{m}|S_{n+1} |1\rangle^m |0\rangle^{n-m+1}$, we require the matrix be made from the first $m$ columns of $S_{n+1}$ as follows  
\begin{align}
    \Omega_{\vec{m}} = \frac{1}{\sqrt{n+1}}
        \overbrace{\begin{bmatrix}
            1 & \omega^{r_1}_{n+1} & \cdots & \omega^{(m-1)r_1}_{n+1} \\ 
            \vdots & \vdots & \ddots & \vdots  \\ 
            1 & \omega^{r_m}_{n+1} & \cdots & \omega^{(m-1)r_m}_{n+1} 
        \end{bmatrix} }^{m \text{ columns}} 
        \left. \phantom{\begin{bmatrix}
            1 \\ 
            \vdots \\ 
            1 
        \end{bmatrix}\hspace{-2.3em}} \right\}{\scriptstyle m \text{ rows}}. 
\end{align}
Here $\vec{r}=\{r_1,\ldots,r_m\}$ are the row power contribution to the phase, associated with measurement outcomes $\vec{m}=(m_1,\ldots,m_{n+1})$. We have $m_r$ repeats of row $r$, which have a row power contribution of $r-1$. For example, for the measurement outcome $\vec{m}=(1,2,1)$, we have $\vec{r}=\{0,1,1,2\}$. Note the order of the multiset $\vec{r}$ doesn't matter since the permanent is invariant under row permutations. By the definition of how $\vec{r}$ and $\vec{m}$ are related, we also have the useful identity
\begin{align}
    \sum_{j=1}^m r_j = \sum_{k=1}^{n+1} (k-1) m_k. \label{eq:usefulid}
\end{align}
Thus we note using Eq.~\eqref{eq:Aper} we have the case that  
\begin{align}
    p \equiv \langle \vec{m}|S_{n+1} |1\rangle^m |0\rangle^{n-m+1} = \frac{\text{Per}( \Omega_{\vec{m}})}{\sqrt{\prod_{i=1}^{n+1}m_i!}}.  
\end{align}
Similarly, for the probability amplitude $\langle \vec{m}|S_{n+1}|0\rangle |1\rangle^m |0\rangle^{n-m}$ we require the matrix be made from the second to $m+1$ columns of $S_{n+1}$ 
\begin{align}
    \Omega'_{\vec{m}} = \frac{1}{\sqrt{n+1}}
        \overbrace{\begin{bmatrix}
            \omega^{r_1}_{n+1} & \omega^{2r_1}_{n+1} & \cdots & \omega^{m r_1}_{n+1} \\ 
            \vdots & \vdots & \ddots & \vdots  \\ 
            \omega^{r_m}_{n+1} & \omega^{2r_m}_{n+1} & \cdots & \omega^{m r_m}_{n+1} 
        \end{bmatrix} }^{m \text{ columns}} 
        \left. \phantom{\begin{bmatrix}
            1 \\ 
            \vdots \\ 
            1 
        \end{bmatrix}\hspace{-2.3em}} \right\}{\scriptstyle m \text{ rows}}. 
\end{align}
The permanent property in Eq.~\eqref{eq:perrule} holds in general for any sized matrix $U$, where multiplying any row or column by a scalar $a$ changes it's permanent from $\text{Per}(U)$ to $a \text{Per}(U)$. Thus we can show the relationship between the two permanents is  
\begin{align}
    &\text{Per} ( \Omega'_{\vec{m}} ) = \text{Per} \left( \frac{1}{\sqrt{n+1}} \begin{bmatrix}
            \omega^{r_1}_{n+1} & \omega^{2r_1}_{n+1} & \cdots & \omega^{m r_1}_{n+1} \\ 
            \vdots & \vdots & \ddots & \vdots  \\ 
            \omega^{r_m}_{n+1} & \omega^{2r_m}_{n+1} & \cdots & \omega^{m r_m}_{n+1} 
        \end{bmatrix} \right) \nonumber \\ 
    &=  \prod_{i=1}^m \omega_{n+1}^{r_i} \text{Per} \left( \frac{1}{\sqrt{n+1}} \begin{bmatrix}
            1 & \omega^{r_1}_{n+1} & \cdots & \omega^{(m-1)r_1}_{n+1} \\ 
            \vdots & \vdots & \ddots & \vdots  \\ 
            1 & \omega^{r_m}_{n+1} & \cdots & \omega^{(m-1)r_m}_{n+1} 
        \end{bmatrix} \right) \nonumber \\  
    &= \omega^{\sum_{k=1}^{n+1} (k-1)m_k}_{n+1} \text{Per} ( \Omega_{\vec{m}} ). 
\end{align}
where we used the identity in Eq.~\eqref{eq:usefulid} to simplify the phase factor as $\prod_{i=1}^m \omega_{n+1}^{r_i} = \omega_{n+1}^{ \sum_{i=1}^m r_i} = \omega^{\sum_{k=1}^{n+1} (k-1)m_k}_{n+1}$. Hence using Eq.~\eqref{eq:Aper} the two amplitudes are related by a simple phase factor 
\begin{align}
    \langle \vec{m}|S_{n+1}|0\rangle |1\rangle^m |0\rangle^{n-m} &= \frac{\text{Per}( \Omega'_{\vec{m}})}{\sqrt{\prod_{i=1}^{n+1}m_i!}} \nonumber \\ 
    &= \omega^{\sum_{k=1}^{n+1} (k-1)m_k}_{n+1} \frac{\text{Per}( \Omega_{\vec{m}})}{\sqrt{\prod_{i=1}^{n+1}m_i!}} \nonumber \\ 
    &= \omega^{\sum_{k=1}^{n+1} (k-1)m_k}_{n+1} p, 
\end{align}
as needed to be proven.

\section{Generating the Resource State} \label{sec:gen}

The main challenge for implementing our protocol is the required entangled resource state $|g_n\rangle$, as defined in Eq.~\eqref{eq:res}. The smallest $n=1$ size requires
\begin{align}
    |g_1\rangle &= \frac{|1\rangle|0\rangle+g|0\rangle|1\rangle}{\sqrt{1+g^2}},
\end{align}
which can be generated easily enough by sticking a single photon through an unbalanced beam-splitter $S_2(\tau)$ with transmissivity $\tau = g^2/(1+g^2)$. This gives us
\begin{align}
    S_2(\tau) |1\rangle|0\rangle = \sqrt{1-\tau} |1\rangle|0\rangle+\sqrt{\tau}|0\rangle|1\rangle = |g_1\rangle, \label{eq:s2g1}
\end{align}
which makes our $n=1$ protocol equivalent to the single-photon quantum scissor teleampifier protocol~\cite{ralph2009nondeterministic}. However, the resource complexity increases for the $n=2$ size, which requires the entangled state
\begin{align} 
    |g_2\rangle &= \frac{|1\rangle|1\rangle|0\rangle|0\rangle+g|1\rangle|0\rangle|0\rangle|1\rangle+g^2|0\rangle|0\rangle|1\rangle|1\rangle}{\sqrt{1+g^2+g^4}}. 
\end{align}
Even though each term has the same number of photons, it is not apparent how one could generate this entangled state from two single-photons. We will describe two different methods for generating these states via post-selection. Note that we are imagining a situation where this resource state is prepared offline, or before the unknown input state has arrived to our amplifier.

\subsection{Using Gaussian Boson Sampling}

\begin{figure}[t]
    \begin{center}
        \includegraphics[width=\linewidth]{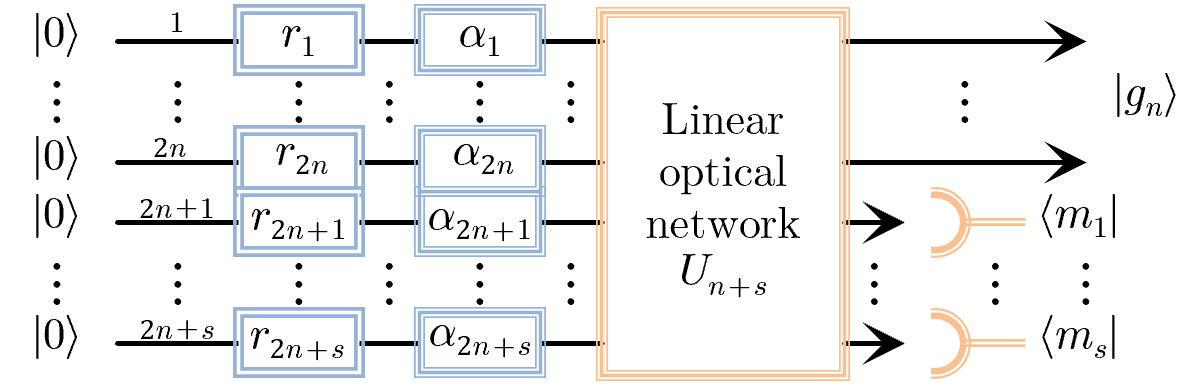}
        \caption{\label{fig:resource-gbs} 
            We can use this device to herald the entanglement resource state $|g_n\rangle$ with high fidelity, if we first optimise over all parameters. The first column are squeezers with $r_j$ squeezing parameters. The second column are displacement operators with $\alpha_j$ displacement parameters. The network $U_{n+s}$ is decomposed into beam-splitters~\cite{clements2016optimal}, whose transmission and phase parameters are also optimised. The photon detection pattern $\langle m_1|\cdots\langle m_s|$ is chosen beforehand.}
    \end{center}
\end{figure}

There are methods based on Gaussian Boson sampling~\cite{hamilton2017gaussian} which allows us to generate these $|g_n\rangle$ entangled states with high fidelity using squeezed displaced light, a linear optical network, and photon detections~\cite{su2019conversion,quesada2019simulating}. It is known that by post-selecting on a particular measurement on some output modes $\langle m_1|\cdots\langle m_s|$, we can herald a wide variety of quantum states in the remaining output modes~\cite{su2019conversion,quesada2019simulating}. In particular, Ref.~\cite{winnel2022achieving} has used this same procedure to create similar quantum states as $|g_n\rangle$, consisting of multiple single-photons entangled over numerous modes. 

We optimised all the parameters in Fig.~\ref{fig:resource-gbs} for fidelity with $|g_2\rangle$ with $g=\sqrt{2}$ gain, and we herald on the photon measurement $\langle1|^5$ (hence in this case $n=2$ and $s=5$). We optimise all the parameters using a machine learning algorithm called basin hopping, as described in Ref.~\cite{sabapathy2019production}. We were able to find a configuration which makes this resource state with high fidelity $F>0.99$, with success probability $\sim 10^{-8}$. We have provided our code which implements this algorithm in Ref.~\footnote{See \href{https://github.com/JGuanzon/optimal-teleamplifier}{https://github.com/JGuanzon/optimal-teleamplifier} for our code which finds the optimal Gaussian Boson sampling-like device to generate the resource states. This uses the Strawberry Fields python library, which includes Ref.~\cite{killoran2019strawberry,bromley2020applications,bourassa2021fast}}.

\subsection{Using Controlled Beam-Splitters}

\begin{figure}[t]
    \begin{center}
        \includegraphics[width=\linewidth]{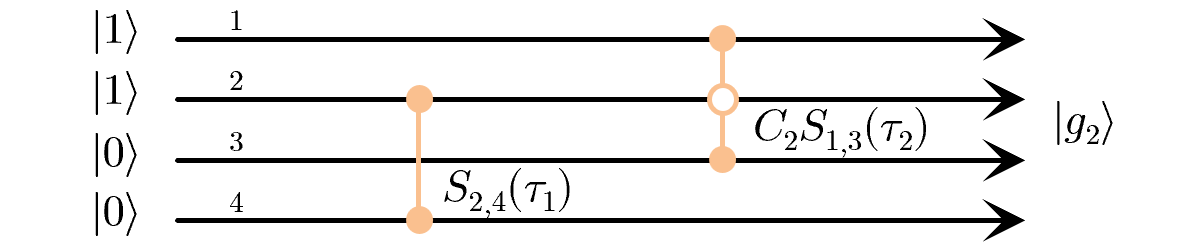}
        \caption{\label{fig:resource-cbs} 
            The protocol for the generation of the resource state $|g_2\rangle$, using two single photons $|1\rangle|1\rangle|0\rangle|0\rangle$, a beam-splitter $S_{2,4}(\tau_1)$ and a photon number controlled beam-splitter $C_2S_{1,3}(\tau_2)$. The vertical bar with solid dots represents a beam-splitter between those modes. The empty dot represents the control mode which implements a beam-splitter between the solid dots if it has zero photons. }
    \end{center}
\end{figure}

We can consider a much more tailored approach for constructing these resource states, using a similar method as Eq.~\eqref{eq:s2g1} where we can generate $|g_1\rangle$ using a single photon and beam-splitter. In Fig.~\ref{fig:resource-cbs}, we show how to generate $|g_2\rangle$ using two single photons, a beam-splitter $S_{2,4}(\tau_1)$ with transmissivity $\tau_1 = (g^2+g^4)/(1+g^2+g^4)$, and a photon-number controlled beam-splitter $C_2S_{1,3}(\tau_2)$ with transmissivity $\tau_2 = g^2/(1+g^2)$. The $C_2S_{1,3}(\tau_2)$ implements the beam-splitter operation only if the control mode contains zero photons, which can be done using linear optical tools with $1/16$ post-selection probability (since it can be done using two non-linear sign gates with 1/4 success probability~\cite{myers2007investigating}). For clarity we include mode numbers as subscripts to the operators and kets, if the modes are unordered.

We will consider how one can use these controlled beam-splitters to construct $|g_n\rangle$ for general sizes $n$. First, recognise that the resource state can be rewritten as 
\begin{align}
    |g_n\rangle &= \frac{1}{\sqrt{\mathcal{N}}}\sum_{j=0}^n (gP)^j |1\rangle^n |0\rangle^n. 
\end{align}
Here $P$ is a mode shifting operator, which simply rotates the position of the states by one in the clockwise direction, i.e. $P|a_1\rangle |a_2\rangle \cdots |a_{2n}\rangle =|a_2\rangle \cdots |a_{2n}\rangle|a_1\rangle$. To better understand the construction method, we can rewrite the last two terms $j\in\{n-1,n\}$ of $|g_n\rangle$ as 
\begin{widetext}
\begin{align}
    |g_{n,M=1}\rangle &= \frac{1}{\sqrt{\mathcal{N}}}\Bigg[ \sum_{j=0}^{n-2} (gP)^{j} |1\rangle^n |0\rangle^n + g^{n-1}|1\rangle|0\rangle^{n-1}|0\rangle|1\rangle^{n-1}
    + g^n|0\rangle|0\rangle^{n-1}|1\rangle|1\rangle^{n-1} \Bigg]  \nonumber \\ 
    &= \frac{1}{\sqrt{\mathcal{N}}}\Bigg[ \sum_{j=0}^{n-2} (gP)^{j} |1\rangle^n |0\rangle^n + g^{n-1}|0\rangle^{n-1}_{2,\ldots,n}|1\rangle^{n-1}_{n+2,..,2n} (|1\rangle|0\rangle + g |0\rangle|1\rangle)_{1,n+1} \Bigg] \nonumber \\
    &= \frac{C_{2}S_{1,n+1}\left(\frac{g^2}{1+g^2}\right)}{\sqrt{\mathcal{N}}}\Bigg[ \sum_{j=0}^{n-2} (gP)^{j} |1\rangle^n |0\rangle^n  + g^{n-1}\sqrt{1+g^{2}}|0\rangle^{n-1}_{2,\ldots,n}|1\rangle^{n-1}_{n+2,..,2n}  |1\rangle_1|0\rangle_{n+1} \Bigg] \nonumber \\ 
    &= \frac{C_{2}S_{1,n+1}\left(\frac{g^2}{1+g^2}\right)}{\sqrt{\mathcal{N}}}\Bigg[ \sum_{j=0}^{n-2} (gP)^{j} |1\rangle^n |0\rangle^n  + g^{n-1}\sqrt{1+g^{2}}|1\rangle|0\rangle^n|1\rangle^{n-1}\Bigg].
\end{align}
\end{widetext}
We can repeat this process multiple times by contracting this remainder term with the next last $j$ term. Hence after using photon number controlled beams-splitters $M$ number of times, $|g_n \rangle$ can be rewritten as
\begin{widetext}
\begin{align}
    |g_{n,M} \rangle &= \frac{ \prod_{i=2}^{M+1} C_{i}S_{i-1,n+i-1}\Big(\frac{ \sum_{k=1}^{i-1} g^{2k} }{ \sum_{k=0}^{i-1} g^{2k}}\Big) }{\sqrt{\mathcal{N}}} \Bigg[ \sum_{j=0}^{n-M-1} (gP)^{j} |1\rangle^n |0\rangle^n  + g^{n-M}\sqrt{\sum_{l=0}^{M} g^{2l}} |1\rangle^M |0\rangle^n |1\rangle^{n-M} \Bigg]. 
\end{align}
\end{widetext}
\begin{widetext}
\begin{align}
    |g_{n,M=n-1} \rangle &= \frac{ \prod_{i=2}^n C_{i}S_{i-1,n+i-1}\Big( \frac{ \sum_{k=1}^{i-1} g^{2k} }{ \sum_{k=0}^{i-1} g^{2k}} \Big) }{\sqrt{\mathcal{N}}} \Bigg[ |1\rangle^n |0\rangle^n  + g\sqrt{\sum_{l=0}^{n-1} g^{2l}} |1\rangle^{n-1} |0\rangle^n |1\rangle \Bigg] \nonumber \\ 
    &= \prod_{i=2}^n C_{i}S_{i-1,n+i-1}\Bigg( \frac{ \sum_{k=1}^{i-1} g^{2k} }{ \sum_{k=0}^{i-1} g^{2k}} \Bigg) S_{n,2n}\Bigg(\frac{ \sum_{k=1}^n g^{2k} }{ \sum_{k=0}^n g^{2k}}\Bigg) |1\rangle^n |0\rangle^n.
\end{align}
\end{widetext}
Note we have used that the normalisation factor is given by $\mathcal{N} = \sum_{j=0}^n g^{2j}$. This representation of $|g_n \rangle$ shows how one can construct it using $n$ single photons, one beam-splitter, and $n-1$ photon number controlled beam-splitters. Hence we have shown that we can in theory construct the required resource states exactly with linear optical tools, with a success probability of $1/16^{n-1}$~\cite{myers2007investigating}. Our proposed method is quite similar to Ref.~\cite{franson2004generation}, however their proposal generates two copies of $|g_n\rangle$ using $2(n-1)$ controlled gates.

\bibliography{paper}

\end{document}